# MemFHE: End-to-End Computing with Fully Homomorphic Encryption in Memory

SARANSH GUPTA, University of California, San Diego, USA
ROSARIO CAMMAROTA, Intel Labs, USA
TAJANA ŠIMUNIĆ ROSING, University of California, San Diego, USA

The increasing amount of data and the growing complexity of problems has resulted in an ever-growing reliance on cloud computing. However, many applications, most notably in healthcare, finance or defense, demand security and privacy which today's solutions cannot fully address. Fully homomorphic encryption (FHE) elevates the bar of today's solutions by adding confidentiality of data during processing. It allows computation on fully encrypted data without the need for decryption, thus fully preserving privacy. To enable processing encrypted data at usable levels of classic security, e.g., 128-bit, the encryption procedure introduces noticeable data size expansion - the ciphertext is much bigger than the native aggregate of native data types. In this paper, we present MemFHE which is the first accelerator of both client and server for the latest Ring-GSW (Gentry, Sahai, and Waters [18]) based homomorphic encryption schemes using Processing In Memory (PIM). PIM alleviates the data movement issues with large FHE encrypted data, while providing in-situ execution and extensive parallelism needed for FHE's polynomial operations. While the client-PIM can homomorphically encrypt and decrypt data, the server-PIM can process homomorphically encrypted data without decryption. MemFHE's server-PIM is pipelined and is designed to provide flexible bootstrapping, allowing two encryption techniques and various FHE security-levels based on the application requirements. We evaluate MemFHE for various security-levels and compare it with state-of-the-art CPU implementations for Ring-GSW based FHE. MemFHE is up to $20k\times$ ($265\times$) faster than CPU (GPU) for FHE arithmetic operations and provides on average $2007\times$ higher throughput than [37] while implementing neural networks with FHE.

## 1 INTRODUCTION

Fully homomorphic encryption (FHE) allows us to apply functions of arbitrary complexity on encrypted data (ciphertext) without the need to decrypt it. This eliminates the need for private key exchanges and decrypting data at the server, raising the bar on security and privacy. This is really critical in areas like healthcare, finance, insurance, etc, which deal with extremely sensitive information but rely on cloud for computing needs [29, 6, 30, 9]. However, computing on encrypted data comes at a huge data and computation cost, resulting in large performance and memory overheads. For example, encrypting an integer in homomorphic domain may explode its size from meagre 4B to more than 20KB. Moreover, homomorphically multiplying two FHE encrypted integers may require 10s of millions of operations. Further, computing with encrypted data may limit the complexity of the function that can be evaluated for a set of encryption parameters. The work in Gentry [17] proposes a procedure, called bootstrapping, to reduce the growth of noise during function evaluation in FHE domain, allowing FHE to perform more complex operations. However, it is extremely expensive and increases the latency of evaluating a homomorphic function by 100-1000×. Recent proposals in [13, 8, 5] make bootstrapping faster and computationally less expensive. Unfortunately, bootstrapping still remains expensive and is the major limiting factor while using FHE to evaluate real workloads. The encryption keys used in such schemes may reach up to GBs in size, adding to the huge capacity and data transfer bottleneck of FHE.

The works in [34, 35, 41, 12, 1, 48] proposed CPU and GPU implementations of RGSW-based FHE schemes [13, 39, 7]. However, they cannot scale enough to provide the speedup needed to

Authors' addresses: Saransh Gupta, sgupta@ucsd.edu, University of California, San Diego, USA; Rosario Cammarota, rosario.cammarota@intel.com, Intel Labs, USA; Tajana Šimunić Rosing, tajana@ucsd.edu, University of California, San Diego, USA.

Table 1. Generations of Fully Homomorphic Encryption

| Gen. | Encrypt-Level | Security | Compute Support | Public Key | Latency/BS | BS Count | Op Accuracy | Type of Apps | Schemes |
|---|---|---|---|---|---|---|---|---|---|
| **1** | Lattice Based | ++ | Add, mul | Largest | 1000x | 1x | - | - | Gentry'09 |
| **2** | Integer | + | Limited predefined ops | Large | 10x | 1x | Approximate | Statistical | CKKS, BGV, B/FV |
| **3+** | Bit/Integer* | +++ | Any arbitrary op | Small | 1x | 10000x | Exact | Any | FHEW, TFHE |

BS: Bootstrapping

make FHE feasible. Most operations in these schemes are based on polynomials and vectors, which are difficult to accelerate due to the limited parallelism and data access provided by current systems. Other hardware-acceleration work in [11, 49, 47, 46] accelerate previous generation schemes which are not truly FHE and support limited functionality. Processing in-memory is an excellent match for the FHE since it provides extensive parallelism, bit-level granularity, and an extensive library of compatible operations which dramatically improving both performance and energy efficiency [31, 14, 15, 27]. It addresses the issue of large data movement by processing data in memory where it is stored. We use Resistive RAM (RRAM) which has low energy requirements, high switching speed, is scalable, and compatible with the CMOS fabrication process.

In this paper, we present the first latest generation end-to-end acceleration of FHE cryptosystem based on [39]. Unlike previous HE proposals, which supported a library of functions, the latest RGSW-based cryptosystem allows computing arbitrary functions on encrypted data. Our proposed MemFHE has two main components, the client and the server PIM accelerators. The client PIM accelerator runs ultra-efficient in-memory operations to not only encode and decode data but also enables ring learning with errors (RLWE) to encrypt and decrypt data. The encrypted data (ciphertext), along with an encrypted version of secret key, are sent to the server PIM accelerator for processing. Server PIM receives the ciphertext from multiple clients and performs operations on ciphertext to generate output. To enable this, server PIM uses PIM-enabled bootstrapping which keeps the accumulated noise low so that the output ciphertext can be decrypted by the intended client. This ciphertext is sent back to the client. In MemFHE, only the client has the means to decrypt the output ciphertext and access the unencrypted data.

To summarize, our specific contributions are:

- We present the first end-to-end acceleration of fully homomorphic encryption in memory. Our design accelerates both the encryption/decryption and the full FHE computation pipelines. MemFHE employs ciphertext-level and operation level parallelism combined with operation-level pipelining to achieve orders of magnitude of performance improvement over the traditional systems. Unlike previous work, we show how PIM can be used to accelerate an application with high data dependency and little data-level parallelism. Our pipelining increases latency by 3% while providing $> 1000\times$ throughput improvement.
- Our server PIM design includes fast bootstrapping, key switching, and modulus switching in memory. It distributes the key memory units to reduce the instances of data contention. It sequentially processes different inputs in different pipeline stages for the best processing throughput.
- We accelerate the bottleneck process of bootstrapping by using a highly pipelined architecture. Our bootstrapping introduces parallel accumulation units, which supports two different types of bootstrapping techniques. We propose a novel implementation for the core bootstrapping operation, Number Theoretic Transform (NTT). Unlike existing works, our NTT doesn't require any special interconnect structure. Moreover, it is flexible and can process many NTT stages without needing extra hardware.

- Our client PIM design includes encryption and decryption. MemFHE enables encryption efficiently in memory by exploiting bit-level access and accelerates dot product with a new in-memory implementation.
- We evaluate MemFHE for various security-levels and compare it with state-of-the-art CPU implementations for Ring-GSW based FHE. MemFHE is up to $20k\times$ ($265\times$) faster than CPU (GPU) for FHE arithmetic operations and provides on average $2007\times$ higher throughput than [37] while implementing neural networks with FHE.

## 2 BACKGROUND AND MOTIVATION

### 2.1 FHE Schemes

Many fully homomorphic encryption schemes have been developed over the past decade. Table 1 summarizes different "generations" of FHE schemes. The first generation include the original design from [17] and its subsequent optimizations. However, they have limited homomorphic capacity due to rapid noise growth during evaluation, restricting the evaluation to few gates at a time. Second generation schemes reduce the noise growth from linear to logarithmic and are based on more standard hardness assumptions. However, they are slow, requiring minutes for simple gate operations (HElib-IBM [26]).

The third generation schemes use weaker hardness assumptions to minimize the bootstrapping time and provide slower noise growth [18, 13, 7]. The work in [39] presented a framework to enable fast bootstrapping for such schemes under different security assumptions. While being the most general, supporting arbitrary functions, allowing many bootstrapping iterations without the need to decrypt, and providing providing control over security-levels, these schemes bootstrap individual boolean gates. They may be slower overall when implementing multi-bit operations. Recent works [22, 56, 37] have shown efficient extension of these schemes for multi-bit operations. Work in this direction promises to deliver faster bootstrapping and better overall application latencies, while providing the ability to perform functions of arbitrary complexity.

### 2.2 FHEW Cryptosystem

FHE in the West (FHEW) cryptosystem [39] is based on the latest generation of FHE schemes, namely FHEW [13] and TFHE [8], and evaluates logic functions on encrypted data, i.e. *ciphertexts*, by evaluating look-up tables (LUTs). This is a foundational work toward realizing the full potential of FHE with more efficient encryption (less data size explosion), and faster bootstrapping for the same level of security as the previous generation schemes. It operates at bit-level, where each data bit is encrypted into pair consisting of a polynomial and an integer using a secret key, $s$, with learning-with-error (LWE) scheme. The encryption is performed for given application parameters, $q$ and $n$, where $n$ is the degree of the polynomial. All operations and data are taken modulus $q$. The typical values of $n$ and $q$, presented in Section 9, results in a bit of data being encrypted into a 0.5-1kb ciphertext. In some cases, FHEW further breaks the ciphertext integers (including each polynomial coefficient) into $d_r$ numbers, each with base $B_r$, to control the growth rate of noise. This further increases the ciphertext size. FHEW operates on LWE-encrypted ciphertexts, utilizing two different encrypted versions of $s$, $EK_B$ and $EK_S$. The encrypted keys may have memory footprint in GBs.

FHEW employs cyclotomic ring-based encryption technique, namely RGSW [18], to operate on the ciphertexts. For each function, like NOR or XOR, that should be applied on the input ciphertexts, FHEW stores a corresponding FHE function in the LUTs. For example, an AND operation between two bits in *plaintext*, translates to simple addition of their corresponding ciphertexts, followed by AND-specific coefficient mapping. This is followed by bootstrapping, which reduces the noise accumulated in the output ciphertext due to function implementation. If not bootstrapped, the output ciphertext may become undecryptable. Most operations in bootstrapping happen over the polynomial part of output ciphertext, using the encrypted version $EK_B$ of $s$. The ciphertext undergoes several *accumulation* iterations during bootstrapping. Bootstrapping works on parameters with similar functionality as that of LWE encryption but have different values, namely $N$, $Q$, $B_g$, and $d_g$. Here, all operations in accumulation happen on integers that have each been decomposed into $d_g$ digits with base $B_g$. The final accumulation output is a pair of polynomials of degree $N$ and modulus $Q$. The final output ciphertext, with reduced noise, is *extracted* out of accumulation result. It is further treated with $EK_S$ encrypted version of $s$ to convert it back to the original LWE-encrypted domain. This process is called key-switching. The key-switched ciphertext is decrypted to get the output.

Apart from the large memory requirements of different FHEW components, the iterative nature and high polynomial degrees of FHEW operations makes it a slow and a memory-intensive process. Most data operations in FHEW are applied over polynomials which have a large compute and memory transfer bottleneck [41]. Efficient polynomial multiplication converts the polynomial into the frequency domain with number theoretic transform (NTT). The digit-decomposed computations of FHEW (i.e. breaking integers into $d_r$ or $d_g$ digits), required back-and-forth polynomial conversions between normal (coefficient) and NTT domain. Cumulatively, these operations make the implementation of FHEW on CPUs/GPUs very slow. Moreover, the huge memory requirement of the third generation FHEW cryptosystem, restricts the development of an effective FPGA/ASIC implementations. In contrast, MemFHE presents the first memory-centric architecture for FHEW cryptosystem. While MemFHE benefits from the large memory density due to its memory-centric approach, processing in memory further enables efficient computations, extreme parallelism, and significantly reduced data movement.

### 2.3 Resistive RAM-based Processing in Memory

Many PIM techniques using RRAM have been proposed recently which implement bitwise operations, arithmetic, and search operations in memory [31, 25, 23, 19, 28], with support for varying bit-widths and data types including binary, integer, fixed point, and floating point. They use the switching-based RRAM processing in memory logic, where operations are governed by the voltage applied at the memory bitlines [31, 23]. The work in [23, 24] implement addition and multiplication using the bitwise operations. A $b$-bit addition is implemented with $b$ serial 1-bit additions, which are further implemented with operations like AND, OR, and XOR. Where, a multiplication operation is implemented by first generating partial products using bitwise AND and then adding them using 1-bit additions. RRAM based PIM may not be completely reliable. However, digital PIM uses single level cells (SLCs) that have been verified to work reliably even with 10-15 % variations in the voltage/resistance [43]. The reliability benefits of SLCs outweigh the added memory-cell requirements for SLCs, e.g. $2\times$ area vs $10^4$ higher endurance [43]. Moreover, FHE parameters consider injection of noise during computation. One can model RRAM's computational unreliability and errors as noise, while selecting FHE parameters.

## 3 RELATED WORK

**FHE on CPUs and GPUs:** The works in [34, 35, 41, 12, 1, 48] proposed CPU and GPU implementations of FHEW [13, 39] and TFHE [7] algorithms. While some implementations optimized the parameters of applications to make it hardware-friendly, others utilized GPU acceleration techniques like memory coalescing and vectorization to improvement the latency of FHEW and TFHE schemes. However, they cannot scale and speedup enough to make FHE feasible.

**FHE FPGA/ASIC Acceleration:** Almost all recent FPGA accelerators for HE are based on the 2nd generation schemes. The work in [11] and [49] perform the basic HE operations for B/FV scheme on FPGAs. The work in [47] implements HE operations for CKKS, another 2nd generation scheme, obtaining significant performance improvements vs. CPU. A recent work in [50] accelerated basic FHE primitives allowing it to essentially support any FHE scheme. However, it does not provide the required memory and compute bandwidth that the latest generation schemes demand. On the contrary, MemFHE implements complete FHE computing pipeline with the latest generation schemes. MemFHE exploits extensive in-memory bandwidth while providing high compute bandwidth by converting 1000s of memory block into computing cores.

**FHE PIM Accelerators:** The work in CiM-HE [46] implements homomorphic arithmetic operations for the 2nd generation B/FV scheme in SRAM. It uses CMOS-based custom memory peripherals to support different operations. While no PIM implementation exists for computing with RGSW schemes, the work in [21] and [45] homomorphically search over data encrypted with a third generation FHE [13]. Their limited functionality restricts practical use.

## 4 MEMFHE SYSTEM OVERVIEW

MemFHE employs an end-to-end privacy-preserving computing system consisting of both client and server implementations. Our architecture is based on the FHEW cryptosystem [39] which provides the slowest noise growth and hence is the most generally applicable class of FHE. MemFHE is implemented completely in memory, using homogeneous crossbar memory arrays and exploits PIM to implement all FHE operations.

All computations in the MemFHE-server happen in encrypted domain. It inputs the encrypted *ciphertexts* and performs the desired operations on the ciphertexts in the basic function unit, $U_{FUNC}$, without decrypting them. Computing in FHE domain leads to the accumulation of noise in the resultant ciphertext. To reduce this noise and keep it below the threshold, server utilizes the MemFHE-bootstrapping. Bootstrapping is the most important but also the slowest process in the MemFHE-server pipeline due to its iterative nature. Hence, we heavily pipeline bootstrapping architecture, so that the slowest operations in bootstrapping happens on different pipeline stages. We introduce novel architectures for various sub-components of bootstrapping and perform operation level optimizations in the bootstrapping core. As a result, MemFHE-server can achieve a high throughput of 170 inputs/ms even for high security parameters, which is 20k× higher than CPU [48].

In addition to the server, we also present MemFHE-client, which provides the input ciphertexts and receives the output of the server. The client is responsible for converting raw data into FHE domain, using a client-specific secret key. The client in FHEW cryptosystem encrypts a bit of data into an LWE ciphertext. MemFHE-client accelerates LWE utilizing efficient in-memory multiply-accumulation and shift operations. The encrypted ciphertext is sent to server along with an encrypted version of the client's secret key. Client also decrypts the output of FHE computation from the server into plaintext form.

## 5 MEMFHE-SERVER ARCHITECTURE

Figure 1 shows an overview of the server's architecture. The goal of MemFHE's server is to provide a high throughput for operations on encrypted data. To achieve this, we create a deep pipeline. As discussed later and evaluated in experiments, bootstrapping is the major bottleneck of the server-side computations. Hence, we use the latency of the slowest bootstrapping stage (i.e. polynomial multiplication) to set the maximum latency of any pipeline-stage in the server. We next present in-memory implementations of all the server components.

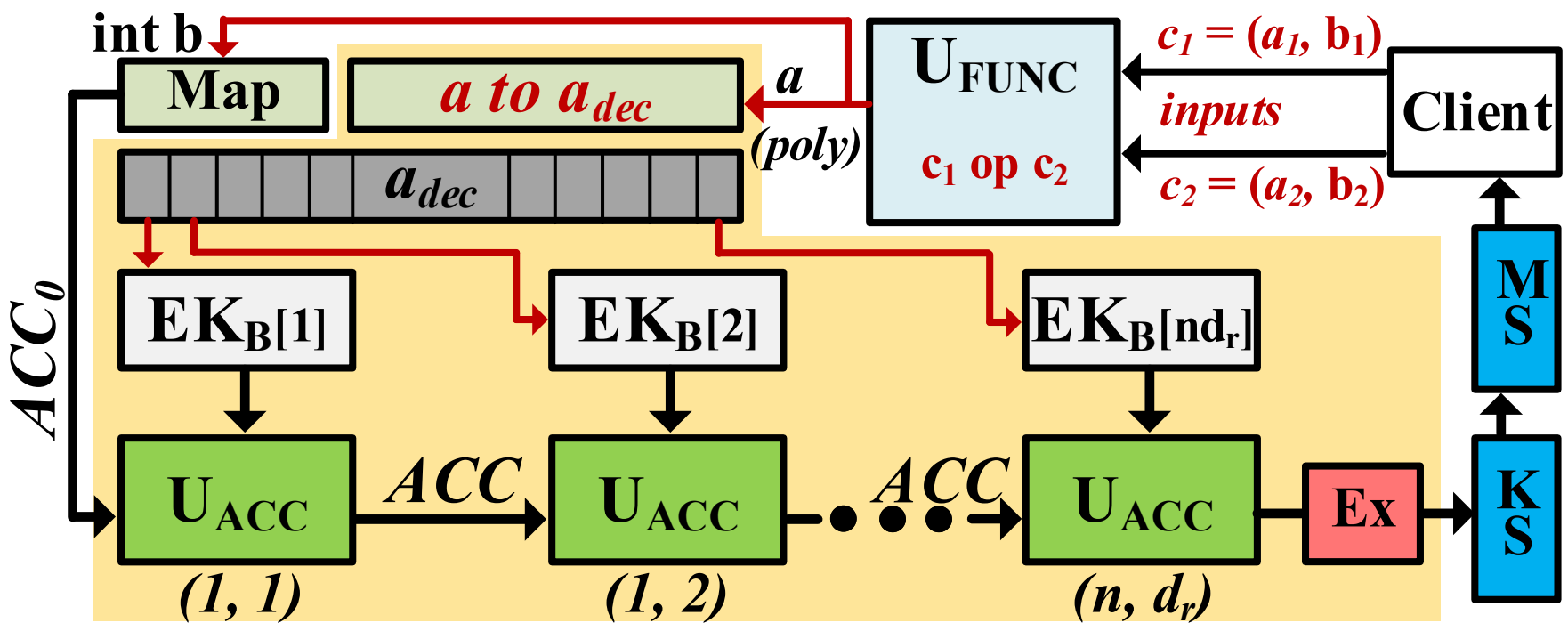

Fig. 1. MemFHE Server Architecture

### 5.1 FHEW Function Implementation

The main strength of FHEW lies in its ability to implement arbitrary functions. FHEW achieves this by translating each boolean function into one or more homomorphic computation steps and then mapping the integer output to a bootstrapping-compatible polynomial, $m_b$. Each element of $m_b$ is set to either $Q/8$ and $-Q/8$, the FHE equivalents of binary '1' and '0'. MemFHE allocates a memory block which stores these translations for all functions. Function implementation is the only process in MemFHE server that follows the client's parameters, $n$ and $q$. FHEW uses polynomial addition, subtraction, and scaling by a constant as computing steps. For example, an AND between two bits is implemented by first homomorphically adding the corresponding ciphertexts (both the polynomial and the integer parts), followed by mapping the integer part of the output ciphertext to $N$-degree polynomial, $m_b$. Then, each coefficient of $m_b$ in $[3q/8,\ 7q/8)$ is set to $Q/8$ and the others are set to $-Q/8$. A complete list of boolean gates and their corresponding FHEW translations are presented in [39]. MemFHE implements computation steps in a memory block, $U_{FUNC}$, executing polynomial additions and subtractions as described in Section 8. Scaling is performed using a series of shift-add operations. Since mapping happens within server's parameters, MemFHE performs it during the initialization stage of bootstrapping discussed in Section 6.1.

### 5.2 Bootstrapping

Implementing functions homomorphically in encrypted domain introduces noise in the ciphertext, which may make it impossible to decrypt the ciphertext. Bootstrapping reduces this accumulated noise. A majority of MemFHE's resources are dedicated to the bootstrapping core. MemFHE transfers the output of $U_{FUNC}$ to bootstrapping. The initialization phase of bootstrapping coverts the output of $U_{FUNC}$ into a server-compatible encryption and initializes a cryptographic accumulator, $ACC$. Then, bootstrapping utilizes a series of accumulation units, $U_{ACC}$, to modify the contents of $ACC$. The accumulation uses $EK_B$ to "decrypt away" the accumulated noise from the output of $U_{FUNC}$. MemFHE supports two types of accumulation schemes, AP [2] and GINX [16]. While GINX is more efficient for binary- and ternary-distributed secret keys, AP is more efficient in other cases [39]. MemFHE chooses the accumulation scheme based on the client's encryption procedure. The output ciphertext with reduced-noise is then extracted from the $ACC$. Section 6 details the implementation of different bootstrapping steps in MemFHE.

### 5.3 Key Switching

Bootstrapping encrypts the output with a different key, $EK_B$ instead of the original key $s$. Key switching is performed to obtain an output encrypted with $s$, so that it can be decrypted by the client. It utilizes the switching key, $EK_S$, which is sent by the client to the server along with the refreshing key, $EK_B$. As shown in [39], key switching uses a base $B_s$ that breaks the integers into $d_s$ digits. The $N$ domain output of $ACC$ gets converted to a client-compatible $n$. Key switching initializes a ciphertext, $c_s$, with an empty polynomial and the integer value of the extracted $ACC$. The ciphertext $c_s$ has the parameters $n$ and $Q$. Each coefficient of the $ACC$ polynomial part, selects elements ($n, Q$ ciphertext) from $EK_S$ and then subtracts them from the existing value of $c_s$. This is repeated for $d_s$ iterations. At the end of each iteration, the $ACC$ polynomial coefficients are divided by the switching base $B_s$.

All operations in key switching are performed modulo $Q$. MemFHE first implements $(d_s - 1)$ divisions as shown in Figure 1. Since $B_s$ is known, MemFHE pre-computes and stores the value of $1/B_s$. Division is now a multiplication with $1/B_s$. To prevent losing data due to rounding errors, the multiplication with $1/B_s$ is performed in full precision, generating twice the number of bits than needed. This happens in parallel for all the coefficients in a row-parallel way. This is followed by a modulo operation with $B_s$. Here we utilize in-memory Montgomery reduction (Section 8) to obtain the modulus of the divided coefficients. Now, we have $N \times (d_s - 1)$ coefficients, that select as many ciphertexts from $EK_S$, and perform sequential ciphertext subtractions. MemFHE employs a tree structure to subtract the ciphertexts. Each computing element of this tree is a memory block. Each blocks perform $x$ sequential subtractions so that the total latency of these subtractions is less than the throughput of the design. Hence, we pipeline the tree stage-by-stage. It takes $\lceil log_2(N.(d_s - 1)/x) \rceil$ tree stages to implement all the subtractions. Each subtraction is followed by Barrett reduction (Section 8 with modulo $Q$. The final output of the tree, $c_s$, is the key-switched output.

### 5.4 Modulus Switching

Lastly, the output of key switching is converted from a modulo $Q$ ciphertext to a modulo $q$ ciphertext. To achieve that, each element is multiplied with $q$ and divided by $Q$ and then rounded off to the nearest integer. MemFHE implements modulus switching in a single memory block. The key-switched ciphertext $c_s$, including its integer part, and is stored vertically in the memory block so that each coefficient is in a separate row. Similar to key switching, MemFHE prestores the value $q/Q$. All the ciphertext coefficients are hence multiplied with $q/Q$ in a row parallel way. Then, a value of 0.5 is added to all the products in parallel using row-parallel addition as detailed in Section 8. Now, for each memory row, the integer part represents the integer nearest to the corresponding coefficient of $c_s.(q/Q)$. We finally take modulus of the output with $q$. Since $q$ is a power of 2 for all security parameters that MemFHE considers, modulo is equivalent to reading $log_2 q$ LSBs of the output. If $q$ is not a power of 2, we use Barrett reduction instead. The output of modulus switching, also the output of server, is a ciphertext with parameter $n$ and $q$, encrypted with secret key, $s$ of the client.

## 6 MEMFHE BOOTSTRAPPING

Bootstrapping inputs an encrypted version of the private key, $EK_B$, also called the refreshing key, along with a ciphertext. The output is a ciphertext corresponding to the input ciphertext but with reduced noise. Bootstrapping performs iterative computations on a cryptographic accumulator, $ACC$. The process involves first *initializing* $ACC$ with the input ciphertext, then implementing an iterative *accumulation* over $ACC$. Each accumulation involves a series of multiplication and addition operations over polynomials. Finally, an element of the final $ACC$ is *extracted* to obtain the output ciphertext. In this section, we discuss the implementation of each of these steps in MemFHE.

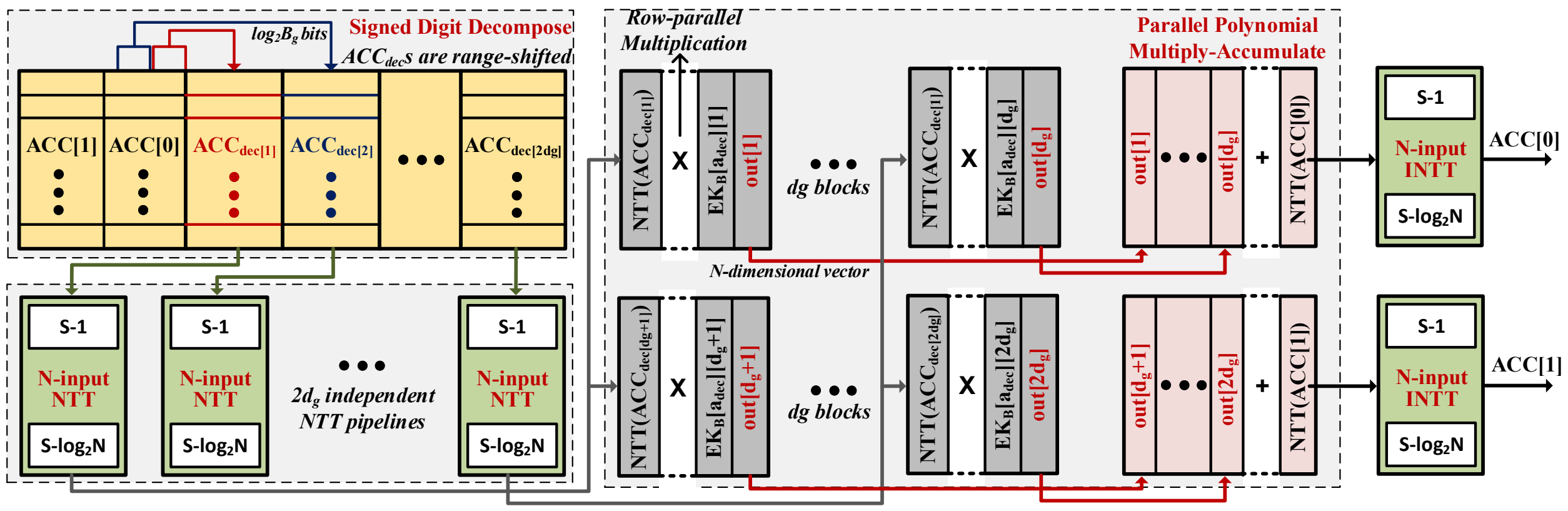

Fig. 2. Accumulation Unit $U_{ACC}$ of MemFHE

## 6.1 Initialization

The initialization phase performs two tasks (i) setting the initial value of $ACC$ and (ii) ensuring that the input ciphertext's polynomial is compatible with the decomposed refreshing key.

**Initializing ACC:** MemFHE performs the mapping discussed in Section 5.1 in this phase. The coefficients of the bootstrapping-compatible polynomial, $m_b$ are each mapped to $Q/8$ and $-Q/8$ based on whether they lie inside or outside an operation-dependent range $(lb, ub)$, $[3q/8,\ 7q/8)$ in the case of AND. To implement this mapping operation in parallel for all the coefficients of $m_b$, we utilize search-based PIM operations. Using exact bitwise-search operations, MemFHE implements in-memory compare operation, which can search a set of memory columns for all the numbers greater, equal, or less than the query. The details of the operation are presented in Section 8. First MemFHE inputs $lb$ as a query and searches for all the numbers greater than $lb$. Then, MemFHE performs searches for the numbers less than $ub$. The final filtered-out rows are initialized to $Q/8$, while the remaining rows are initialized to $-Q/8$. The resultant $m_b$ is the initial $ACC$ value.

**Polynomial's Compatibility with $\mathbf{EK_B}$:** The input ciphertext's polynomial $a$, needs to be made compatible with the decomposed refreshing key, $EK_B$. The polynomial $a$ undergoes the same set of operations as those discussed in key switching, except for subtractions, with parameters $n$, $B_r$, and $d_r$ instead of $N$, $B_s$, and $d_s$. It results in $n \times d_r$ coefficients for each input. We call them $a_{dec}$. For the bootstrapping pipeline to work, all of the $n \times d_r$ $U_{ACC}$ units should receive elements from $a_{dec}$s belonging to different inputs. Hence, we introduce an $n \times d_r$-sized register, in which word$_i$ is fed directly to $U_{ACC-i}$.

## 6.2 Accumulation

The inputs to the accumulation function include the decomposed representation of $a$ ($a_{dec}$ from the initialization step, an RGSW encrypted refreshing key, $EK_B$, and the output of initialization step, a pair of polynomials of degree N. Accumulation preforms iterative multiplication of this key with $ACC$ and then addition back to $ACC$. It is the slowest part of bootstrapping due to high data dependency between the iterations. It adds the result of multiplication in each iteration to the accumulator. The dependency of the input of one ciphertext element on the output of the previous one further prohibits the functions from being parallelized across the ciphertext elements. However, each ciphertext element is a high-degree polynomial, allowing parallelize over the polynomial length.

*6.2.1 AP Bootstrapping:* Traditionally, refreshing key is an $n$-dimensional vector where each element of the vector is either an $N$-degree polynomial or a pair of those. However, in AP bootstrapping instead of each element of $EK_B$ being an $N$-degree polynomial, it is a pair of $2d_g$ polynomials of degree $N$. Each dimension of the vector is further represented using the pair ($B_r$, $d_r$). Hence, the AP refreshing key is a three dimension matrix where each element of the matrix is a pair of $2d_g$ N-degree polynomials. MemFHE stores the refreshing key in $n \times d_r$ memory blocks such that each block stores $2B_r.d_g$ polynomials. Each $EK_B$ memory block is assigned to the corresponding accumulation unit. The main computation of the AP bootstrapping is to perform accumulation function on ACC $n \times d_r$ times. Each step involves a multiplication of the current ACC value with an element of $EK_B$ as $ACC \leftarrow ACC \diamond EK_B$.

**Accumulation Unit ($U_{ACC}$):** We design a bootstrapping pipeline such that the accumulation logic consists of $n \times d_r$ accumulation units, $U_{ACC}$. The unit address $(i, j)$, where $0 \leq i < n$ and $0 \leq j < d_r$, corresponds to the $(i \times d_r + j)$th accumulation iteration. While the units cannot operate on multiple iterations of a single ciphertext in parallel, they can process different ciphertexts in a pipelined fashion. Each unit receives the corresponding value from $a_{dec}$ memory and uses it to select an element from $EK_B$ for multiplication. Since all units input $EK_B$ in each iteration, it introduces a fetch bottleneck at the $EK_B$. To reduce this problem, $EK_B$ is split over multiple memory blocks, with each $U_{ACC}$ having a local $EK_B$ memory. $EK_B$ is independent of inputs and populated once.

Since FHEW is based on RGSW encryption scheme, the multiplication in the accumulation stage happens on digit-decomposed operands to reduce the growth of noise. As explained later, the SDD tile in $U_{ACC}$ performs digit decomposition on the two $N$-degree polynomials of ACC, splitting each coefficient of ACC into $d_g$ numbers with $log_2 B_g$ bits each. $EK_B$ is already digit-decomposed. The output of SDD tile, digit-decomposed $ACC_{dec}$, contains $2d_g$ polynomials of degree $N$, similar to each part of $EK_B$ pair polynomials. Now $U_{ACC}$ performs $4d_g$ polynomial-wise multiplications in parallel, $2d_g$ between $ACC_{dec}$ and each part of the $EK_B$ pair as shown in Figure 2. To make the multiplication efficient, all the polynomials are converted in NTT domain before multiplying. $U_{ACC}$ employs $2d_g$ NTT pipelines and converts $ACC_{dec}$ into NTT domain. The details of our NTT pipeline are presented in Section 6.2.3. $EK_B$ is already in NTT domain. Polynomials in NTT domain are stored in a row-parallel way, such that each coefficient is stored in a separate row as shown in Figure 2. Then, we perform row-parallel multiplication between the polynomials. After multiplication, all products are accumulated to generate a pair of polynomials that serve as the output ACC. Before sending the output to the next unit, $U_{ACC}$ converts it back to the coefficient (non-NTT).

**Signed Digit Decompose (SDD):** Signed digit decompose (SDD) decomposes a pair of polynomials into multiple polynomials. The core operation is to break each polynomial coefficient (originally $log_2 Q$ bits) into smaller $log_2 B_g$ bit signed numbers. As shown in Table 2, $B_g$ is always a power of 2, making the process simpler. SDD consists of one or more memory blocks which perform iterative modulus-division operations, as shown in Figure 2. In each iteration, MemFHE selects $log_2 B_g$ LSBs (remainder of the division by $B_g$) from the coefficients, preserving the remaining bits (quotient of the division). The selected LSBs represent the first $log_2 B_g$-bit number. This process is repeated $d_g$ times, decomposing all coefficients into into $d_g$ $log_2 B_g$-bit numbers. Hence, in the beginning of each iteration, we first change the range of the coefficients from $[0, Q)$ to $[-Q/2, Q/2]$ by subtracting $Q$ from all inputs in $[Q/2, Q)$, mapping them to $[-Q/2, 0)$. MemFHE implements this operation in parallel for all the coefficients of the input polynomial. Coefficients are stored in different rows, occupying the same set of memory columns. We search for all numbers greater than $Q/2$ using MemFHE's in-memory parallel compare operation discussed in Section 8. MemFHE then subtracts $Q$ from all the filtered coefficients. Similarly, the selected LSBs (remainders) are sign-extended, where MemFHE copies the $(log_2 B_g - 1)$th bit for all the coefficients in parallel. Then, all negative remainders are made positive. MemFHE achieves this by searching the MSB bits of all the remainders in parallel (one remainder per coefficient per iteration) and subtracting $Q$ from the filtered remainders.

*6.2.2 GINX Bootstrapping:* The decision to run either AP or GINX bootstrapping is based on the type of secret key used by the client. As shown in [39], GINX works better in case of binary and ternary secret keys, while AP works better for other. GINX bootstrapping differs from AP in two major ways. First, it utilizes binary secret keys, resulting in a smaller refreshing key $EK_B$. $EK_B$ in GINX has a dimension of $n \times 2$, instead of AP's $n \times B_r \times d_r$. Each element consists of $2d_g$ polynomials of degree $N$, the same as AP. Second, the bootstrapping function in GINX involves extra multiplicative and additive terms to generate the effect of input-dependent polynomial rotation. Specifically, the bootstrapping follows:

$$ACC \leftarrow ACC + (X^m - 1)(ACC \diamond EK_B),$$

where $m = \lfloor a(i) \times (2N/q) \rfloor$ for $i$th coefficient of the input ciphertext polynomial $a$. $(X^m - 1)$ is a monomial representing GINX's "blind rotation" by $m$. This encodes the input in the form of the powers of polynomial. The state-of-the-art implementation PALISADE [48] pre-computes $(X^m - 1)$ for all possible values of $0 \leq m < 2N$ and maintains a library of their NTT counterparts. Based on the $m$ corresponding to a $U_{ACC}$, PALISADE selects a value from the library and then multiply it with $U_{ACC}$'s output. This creates a data transfer bottleneck in a pipelined architecture like MemFHE's, where many units need to access the library simultaneously. On the contrary, MemFHE exploits the bit-level access provided by PIM to implement this "rotation" efficiently.

MemFHE uses the same architecture to implement GINX as that for AP. GINX requires $n \times 2$ $U_{ACC}$ units. Here, unlike AP, $EK_B$ input to $U_{ACC}$ is independent of the polynomial part $a$ of the ciphertext. Like in the case of AP, the SDD tile of $U_{ACC}$ first decomposes input $ACC$, $U_{ACC}$ then performs the same polynomial-wise multiplication and subsequent addition, and finally converts them to coefficient domain using INTT. Now, the output of addition represents $prod = (ACC \diamond EK_B)$ in coefficient domain. We now perform in-memory row-parallel rotation on $prod$ as discussed in Section 8. MemFHE finally adds the rotated $prod$, $prod_r$, to pre-decomposed $ACC$ and finally subtracts $prod$. The output is the GINX accumulated $ACC$ in coefficient domain.

*6.2.3 NTT and INTT Pipeline.* Number theoretic transform (NTT) is a generalization of fast Fourier transform (FFT) that performs transformation over a ring instead of complex numbers. In FHE, it is mainly used in polynomial multiplication where it converts a polynomial (by default in coefficient domain) into its frequency (NTT) domain equivalent. A polynomial multiplication in coefficient domain translates to an element-wise multiplication in NTT domain, enabling extensive parallelism for high-degree polynomials. However, the process of converting to and from NTT domain is complex. The state-of-the-art implementations of NTT [47, 42] utilize algorithms where the coefficient access pattern for an $n$-degree polynomial changes for each of the $log_2 n$ stages of NTT pipeline. Instead, we utilize Singleton's FFT algorithm proposed in [51] and later accelerated in [36, 54, 10] to implement MemFHE's NTT pipeline. Figure 3a shows the signal flow graph for Singleton's FFT algorithm. We observe that the coefficient access pattern for the algorithm remains the same for every stage. MemFHE exploits this property to avoid using NTT-specific interconnects.

**Data Mapping:** Figure 3b shows the data layout of one NTT stage in MemFHE. We write an $n$-degree input polynomial, $a$, in $n/2$ rows such that a pair of coefficients with indices $2i$ and $(2i+1)$ share the $i$th row of the memory block. All such pairs are hence written in separate rows, utilizing the same columns. A twiddle factor is associated with each pair, which is pre-computed and stored in the corresponding row. Each pair generates the $i$th and $(i + n/2)$th coefficients of the output polynomial in $i$th row of the block.

**Computation:** Each NTT stage of MemFHE performs three compute operations. First, we perform row-parallel multiplication between the coefficients with odd indices $(2i + 1)$ and the corresponding twiddle factor $W$. Second, we add the generated products to the coefficients with even indices ($2i$) in a row-parallel way to generate the first $n/2$ coefficients of the output polynomial. Lastly, we subtract the products from the even-indexed coefficients in a row-parallel way to obtain the remaining output coefficients. The details of the row-parallel operation execution are presented in Section 8.

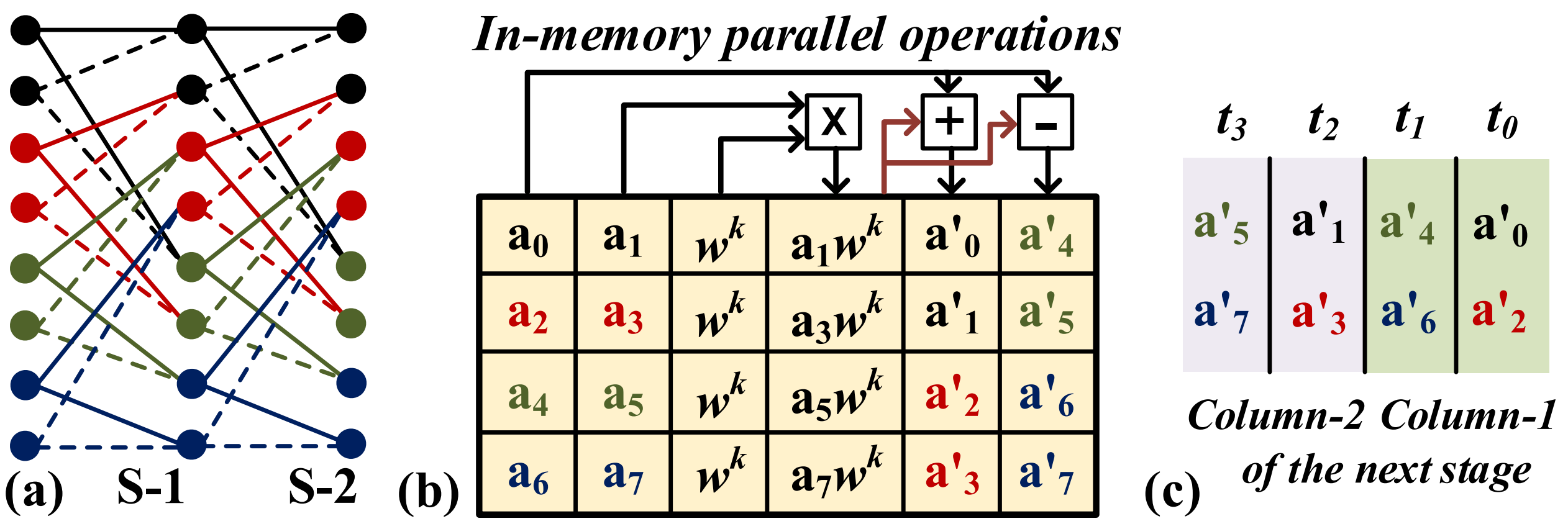

Fig. 3. Singleton's NTT in MemFHE

**Stage-to-Stage Data Transfer:** Figure 3c shows the data transferred in each transfer phase. We perform column-wise data transfer, where each column consists of one bit from all (or a subset of) rows of the memory block. In one data transfer phase, $q$ column transfers can transfer as many $q$-bit numbers as the rows in the memory. As discussed in data mapping, the output polynomial is present in $n/2$ rows such that indices $[0, n/2-1]$ are stored in one set of columns and the remaining indices in the another set of columns. Hence, we need four data transfer phases. The first data transfer reads the even-indexed coefficients from $[0, n/2-1]$ and write them to the next stage according to the data mapping scheme, while the second data transfer does the same for the even-indexed coefficients from $[n/2, n-1]$. Similarly, third and fourth data transfer phases deal with odd-indexed coefficients. These data transfers read selected rows from one memory block, send it over a conventional local interconnect, and write them at a contiguous location of the destination memory.

**Operation Pipeline:** We pipeline our NTT implementation at the granularity of an NTT stage. Hence, the pipeline depth is given by the number of NTT stages: $(n \times dr) \times (2log_2 n + 2)$. Each stage works in parallel over different inputs. As discussed in Section 9, each MemFHE memory block contains 1024 rows. Hence, one memory block can implement an NTT stage for up to 2048-degree polynomial, requiring a total of $11(log_2 2048)$ memory block for whole NTT. For $n < 2048$, we perform NTT over $m = 2048/n$ inputs at the same time in parallel, while requiring only $log_2 n$ stages in the pipeline. In order to maintain the computation and data transfer characteristics, we interleave the inputs as shown in Figure 3e. Here, the output throughput of the pipeline becomes $m\times$ the original throughput. For $n > 2048$, MemFHE allocates multiple memory blocks per stage and implements a deeper pipeline. Since MemFHE's NTT is stage-wise pipelined, the throughput of the larger NTT is the same as that for $n = 2048$.

**Inverse NTT (INTT):** NTT and INTT utilize the same hardware and have identical data-mapping, computation, transfer, and pipelining schemes. The two operations differ only in the twiddle factors they use. During pre-compute step, INTT pipeline generates the twiddle factors, $w^{-k}$, which are inverse of those used in NTT. The rest of the process is the same.

### 6.3 Extraction

After accumulation, $ACC$ consists of a pair of polynomials. Extraction is a simple mapping process that converts $ACC$ to a ciphertext. The first polynomial of $ACC$ represents the polynomial part of the bootstrapped output ciphertext. Whereas, the constant term (corresponding to degree-0) of the second polynomial represents the integer part. To reverse the mapping operation that occurred during initialization phase, $Q/8$ is added (modulo $Q$) to the integer part.

## 7 MEMFHE CLIENT ARCHITECTURE

### 7.1 Encryption

Client encryption converts a message bit, $m$, into a ciphertext of the type $(a, b)$, where $a$ is an integer polynomial of length $n$, while $b$ is an integer. This encryption utilizes learning with errors (LWE) encryption technique [38, 44, 39] and is defined as $LWE_s(m) = (a, b) = (a, (a.s + e + m')\ mod\ q)$, where $m'$ is an encoded version of $m$, $s$ is the secret key, and $e$ is an integer error added to the message.

Evaluating $m'$ involves dividing the message, $m$, with a message modulus $t$ and then multiplying the output with the application parameter, $q/2$. According to the state-of-the-art implementation in [48] and the security parameters presented in [39] and Section 9, $t$ and $q$ are always powers of 2. Hence, MemFHE scales $m$ to $m'$ using in-memory shift and add operations. We first extract the $log_2 t$ LSBs of m. Then, in-memory multiplication with $q/2$ is simply a left shift operation on $m\%t$ by $log_2(q/2)$. Since all the operations in encryption are done modulo $q$, we extract the $log_2 q$ LSBs of the output. In the case when $q$ is not a power of 2, we perform modulo operations as described in Section 8.

Generating integer $b$ requires a dot product between vectors $a$ and $s$, followed by adding $e$ and $m'$. To generate this dot product, we utilize the secret key memory, $SK_{mem}$. It stores the vector corresponding to secret key $s$ in a row-parallel way such that all the elements of $s$ occupy the same set of memory bitlines and each element is stored in a different row. The incoming vector $a$ is written such that the corresponding elements of $a$ and $s$ are present in the same row.

We implement row-parallel integer multiplication between the elements of the two vectors. Our row-parallel execution performs vector-wide multiplication with the same latency as that of a single multiplication, discussed in Section 8. This is followed by an addition of all the products. To add, we perform column parallel in-memory addition operations on the output products such as those proposed in [14] but using the in-memory switching techniques instead of sense amplifier based operations of [14]. In the following discussion, we denote the bitwidth of each product (i.e. $log_2 q$) with the letter $p$. Here, we accumulate each bit position independently, so that $k$ $p$-bit numbers are reduced to $p$ $log_2 k$-bit numbers after $(k-2)$ column parallel 1-bit additions for each of the $p$ bit position. To further reduce the output to a single number, we transpose the output of column-parallel addition so that the outputs for all $p$ columns are stored in the same row. It takes $p$ data transfers, $log_2 k$ bits per transfer, to read the outputs column-wise and store them in a row. We then perform bit-serial addition to obtain the final integer output, which takes $p \times log_2 k$ 1-bit additions. This output represents the dot product $a.s$, to which we add integers $e$ and $m'$.

### 7.2 Decryption

Client decryption converts the server's output ciphertext, $(a, b)$, back to a bit message, $m$, as $Round(4/q * (b - a.s))$, where $s$ is the client's private key. MemFHE first uses the dot product implementation of MemFHE's encryption to obtain $a.s$, followed by a subtraction operation with $b$. The subtraction is followed by a modulo $q$ operation, where MemFHE simply reads the $log_2 q$ LSBs of the output. Scaling is done with $4/q$ by discarding the $log_2(q/4)$ LSBs. $Round(.)$ is implemented similar to the rounding function discussed during modulus switching in Section 5.4.

Table 2. MemFHE Security Parameters [39]

| Set | Security | n | q | N | $log_2 Q$ | $B_s$ | $B_g$ | $B_r$ |
|---|---|---|---|---|---|---|---|---|
| **Classical** | | | | | | | | |
| **STD128** | 128-bit | 512 | 512 | 1024 | 27 | 25 | $2^7$ | 23 |
| **STD192** | 192-bit | 512 | 512 | 2048 | 37 | 25 | $2^{13}$ | 23 |
| **STD256** | 256-bit | 1024 | 1024 | 2048 | 29 | 25 | $2^{10}$ | 32 |
| **Quantum – Safe** | | | | | | | | |
| **STD128Q** | 128-bit | 512 | 512 | 2048 | 50 | 25 | $2^{25}$ | 23 |
| **STD192Q** | 192-bit | 1024 | 1024 | 2048 | 35 | 25 | $2^{12}$ | 32 |
| **STD256Q** | 256-bit | 1024 | 1024 | 2048 | 27 | 25 | $2^7$ | 32 |

## 8 MEMFHE COMPUTATIONS

Here, we detail PIM implementation of MemFHE operations.

**Vectorized Data Organization:** MemFHE implements vectorized-versions of its operations. An input vector, with $n$ $b$-bit elements, is stored such that $n$ elements occupy $n$ different rows with but share the same $b$ memory columns.

**Row-parallel Addition and Multiplication:** A $b$-bit addition in MemFHE is implemented using bitwise AND, OR, and XOR and requires $(6b+1)$ memory cycles [23]. Similarly, multiplication is performed by generating partial products and serially adding them. MemFHE optimizes the multiplication in [24] by sharing the memory cells among intermediate outputs of addition and utilizing faster operations proposed in [23]. This significantly reduces the time to perform full precision $b$-bit multiplication from $(13b^2 - 14b - 6)$ to $(7b^2 + 4b)$ memory cycles, while the total memory required reduces from $(20b - 5)$ to $13b$. This increase the maximum possible multiplication bitwidth from 51 bits in [24] to 78 bits in MemFHE.

**Modulus/Modulo:** Modulus operation gives the remainder of a division. In the context of FHE, modulus is used to avoid overflow during computation. Hence, most operations in MemFHE are followed by modulus. In most cases in MemFHE-server, modulus is taken with respect to a prime number. We perform PIM variants of Barrett [3] (for addition) and Montgomery [40] (for multiplication) reductions using shift and add operations, as done in [42]. This requires prior knowledge of the modulus base, which is governed by the security parameters (and hence known) in MemFHE. If taken with respect to a power of 2, then modulus just selects the corresponding LSBs of the input.

**Comparison:** Comparison operation in MemFHE can compare an input query with the data stored in MemFHE's memory blocks. We exploit the associative operations proposed in [19] to search for a bit of data in a memory column. To compare data stored in $b$ columns and $r$ rows of a memory block with a $b$-bit query, we perform bit-by-bit search. Starting from MSB, associative search is applied for each memory column and all memory rows. Associative search circuit [19] selects all rows where there is a mismatch between the stored and query bit.

**Rotation:** Rotation in MemFHE is equivalent to reading out a memory row (column), bit-wise rotating them at the input register of the block and writing it back.

**Shift:** MemFHE implements shift operation by simply selecting or deselecting bitlines for the corresponding LSB/MSBs. If sign-extension is required, then MemFHE copies the data stored at the original MSB bitline.

# 9 EVALUATION

## 9.1 Simulation Setup

We simulate MemFHE using a cycle-accurate simulator. The simulator considers the memory block size ($1024 \times 1024$ bits in our experiments), the precision for each operation, the degree of polynomials, the locations and the organization of the data. We use HSPICE for circuit-level simulations and calculate energy consumption and performance of all the MemFHE operations with 28nm process node. We adopt an RRAM device with VTEAM model [32] and switching delay of 1.1ns [52]. The parameters of the model have been set to mimic the behavior of practical RRAM memory chips [53]. RRAM components of the design have a SET and RESET voltage of 2V and 1V respectively, with a high-to-low resistance ratio of $10M\Omega/10k\Omega$. A detailed list of parameters is presented in [31, 27]. However, the proposed architecture works with most processing in memory implementations based on digital data.

MemFHE is based on the FHEW cryptosystem of PALISADE library [48]. We perform our evaluation over multiple security parameter sets as described in [39] and summarized in Table 2.

## 9.2 MemFHE-Server Pipeline Analysis

Figure 4 shows the throughput, latency, energy consumed, and memory required for one MemFHE-server pipeline with different parameter settings. We compare the throughput-optimized and area-optimized implementations of the pipeline. The two implementations differ in the way they pipeline NTT/INTT. While the area-optimized version follows the stage-wise pipelining mechanism discussed in Section 6.2.3, the throughput-optimized design implements a finer-grained pipeline. It further breaks an NTT stage into three pipeline stages, first for multiplication with twiddle, second for reduction of the product and addition/subtraction, and the third for final reduction and data transfer to the next stage.

**Throughput-Optimized MemFHE:** We observe that the four design metrics change significantly with the security levels. Throughput is highly dependent on $Q$, the bitwidth of server-side computations. More precisely, throughput varies approximately with $(log_2Q)^2$. This happens because the slowest operation of the pipeline, i.e. the coefficient-wise multiplication, has an implementation latency of $O(Q^2)$ in MemFHE. MemFHE's latency is dependent on $Q^2$ as well as the polynomial degree of input ciphertext, $n$, and parameter $d_r$ and varies approximately with $n.d_r.(log_2Q)^2$. MemFHE-server consumes a total energy of 34 mJ (164 mJ) for processing an input in 128-bit classical (quantum-safe) FHE setting. While the quantum-safe implementations consume higher energy than their classical counterparts, the difference reduces as the security-level increases. The total memory consumed by MemFHE's server changes with different parameter settings as well. It varies approximately with $n.N.d_g$, consuming 37 GB (47 GB) for a complete server pipeline running 128-bit classical (quantum-safe) FHE. We further observe that the accumulation of cryptographic accumulator, $ACC$, consumes on average 96.5% of the total memory requirement of the server pipeline, while contributing 99.7% to the total latency. Accumulation makes up 99.9% of the total bootstrapping computational effort. Hence, this effectively represents the performance of bootstrapping.

**Area-Optimized MemFHE:** While MemFHE provides extensive throughput benefits, it takes considerable amount of area. Moreover, since memory is the main resource in MemFHE, we optimized our implementation for area. We observe that an area-optimized MemFHE-server pipeline consumes 2.5× less memory resources on average as compared to the throughput-optimized design, while reducing the throughput by approximately 2.2×. In contrast, the latency increases by 75%. This happens because we reduce the number of pipeline stages by 3× in the area-optimized design but at the same time increase the latency of each pipeline stage by 2.2×. Since the operations remain the remain in both the designs, their total energy consumption is similar. This highlights one of the advantages of PIM as pipelining doesn't have operational and storage overhead since outputs of most operations are generated in the memory block and hence stored inherently.

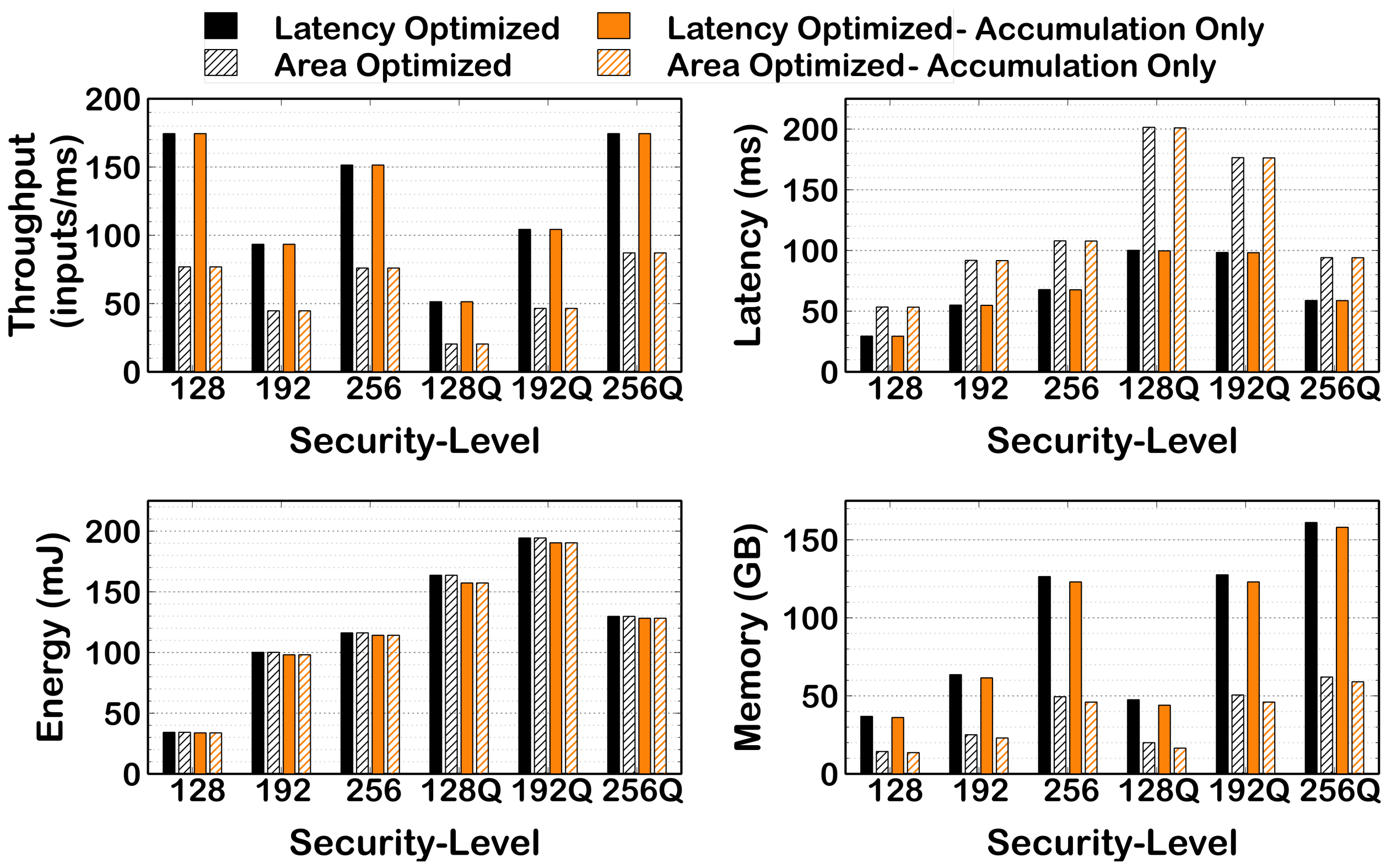

Fig. 4. MemFHE-server pipeline results for a bitwise operation. The suffix Q represents quantum-safe security guarantee.

Table 3. MemFHE Key Sizes (in MB)

| | STD128 | STD192 | STD256 | STD128Q | STD192Q | STD256Q |
|---|---|---|---|---|---|---|
| $EK_S$ | 253 | 925 | 1269 | 1719 | 1750 | 1013 |
| $EK_B$ (AP) | 322 | 897 | 1920 | 1150 | 2304 | 1792 |
| $EK_B$ (GINX) | 14 | 39 | 60 | 50 | 72 | 56 |
| Total (AP) | 575 | 1822 | 3189 | 2869 | 4054 | 2805 |
| Total (GINX) | 267 | 964 | 1329 | 1769 | 1822 | 1069 |

### 9.3 MemFHE-Server Scalability

We take the area-optimized MemFHE for different security-levels and scale it to the given memory size. MemFHE has a minimum memory requirement, which is storage needed for the refreshing and switching keys. The different key sizes in MemFHE are presented in Table 3. To scale down from a pipeline's ideal memory size described in Section 9.2 and Figure 4, we reduce the number of NTT cores. To scale up, we increase the number of parallel pipelines.

Figure 5 shows the throughput of the server for different security levels under different memory constraints. Missing bars in the figure show the cases when the available memory is not sufficient to implement MemFHE. We observe that MemFHE's throughput changes almost linearly with the total memory availability. It increases from the ideal 77 inputs/ms with 14 GB memory consumption to 307 inputs/ms with 64 GB for 128-bit security level, while decrease to 7 inputs/ms with 2 GB memory size. However, in some cases the changes isn't linear. For example, for the quantum-safe 128-bit security configuration, MemFHE's throughput of 20 inputs/ms doesn't change when going from the ideal 20 GB to 32 GB. This happens because the increase in memory is not sufficient to support two pipelines. At the same time, increasing the memory availability further to 64 GB increases the throughput by 3× to 61 inputs/ms because 64 GB memory has enough resources to fit three STD128Q pipelines.

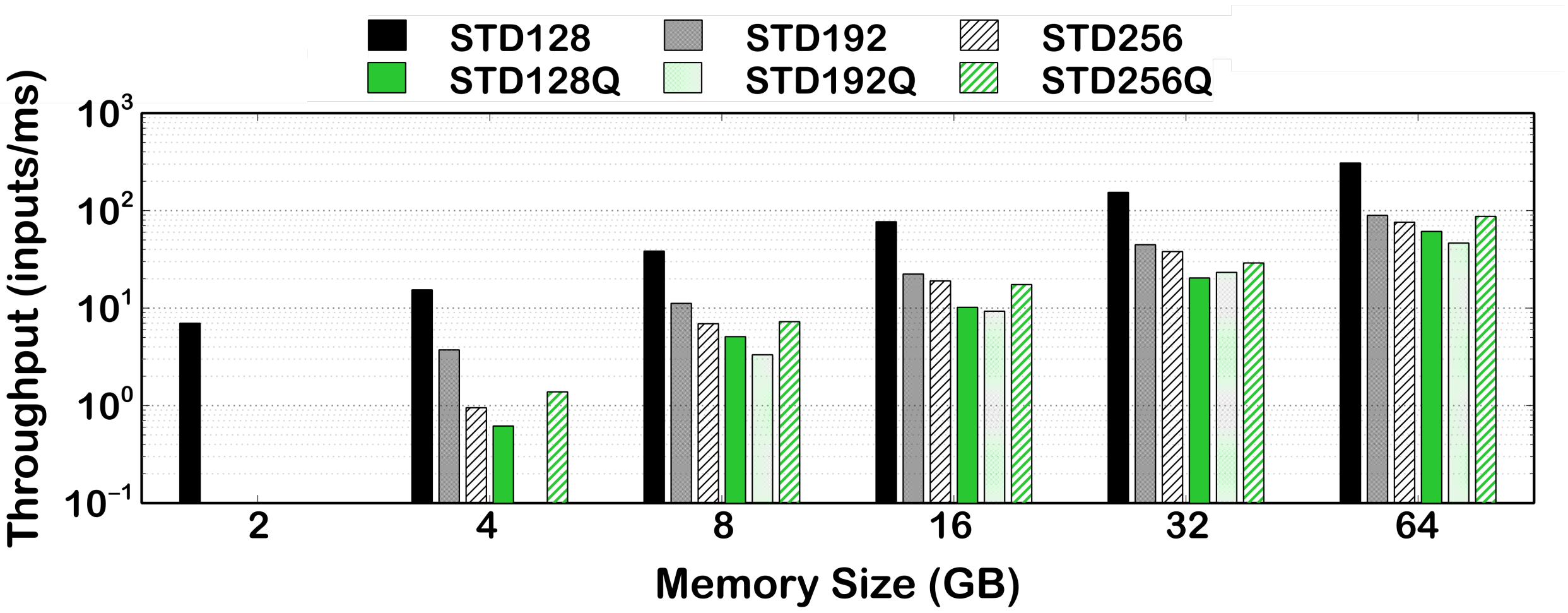

Fig. 5. MemFHE-server throughput for different memory sizes. The missing bars represents memory lower than the minimum required size.

### 9.4 MemFHE Client Analysis

MemFHE-client encrypts bits to ciphertexts and decrypts processed ciphertexts back to bits. Figure 6a shows the encryption latency and energy consumption for MemFHE-client at different security levels for a bit. Decryption involves the same operations and has roughly the same latency as that of encryption. The latency of encryption depends on the ciphertext modulus, $q$, and the polynomial degree, $n$. As expected, the dot product $a.s$ is the slowest operation in encryption, taking 98% of the total latency. Encrypting a bit to a 128-bit (256-bit) quantum-safe ciphertext takes 3 us (5.5 us), while it consumes 4 nJ (9.8 nJ) of energy.

MemFHE requires a total of 128 KB (256 KB) memory (one memory block) for generating a 128-bit (256-bit) quantum-safe ciphertext. However, similar to MemFHE-server, the client is also scalable and employs multiple encrypting-decrypting memory blocks for processing multiple inputs in parallel. Figure 6b shows how the throughput of the MemFHE-client changes with the available memory sizes. The figure shows the combined encrypt-decrypt throughput. Each memory block in MemFHE can be dynamically configured to run either encryption or decryption. We observe that the client's throughput increases linearly with the increase in the total memory size, going from 0.2 inputs/us for 256 KB memory to nearly 47 inputs/us for 64 MB for quantum-safe 256-bit encryption.

### 9.5 Arithmetic Operations in MemFHE

In this subsection, we show the end-to-end performance of MemFHE while implementing addition and multiplication. We utilize Kogge-Stone adders for addition operation as well as accumulation of partial products during multiplication. This reduces the critical path of the circuits and hence, the end-to-end latency for an input. Provided sufficient independent inputs, MemFHE can implement all these operations with the same throughput as shown in Section 9.2, processing up to 174 inputs/ms at 256-bit quantum-safe security.

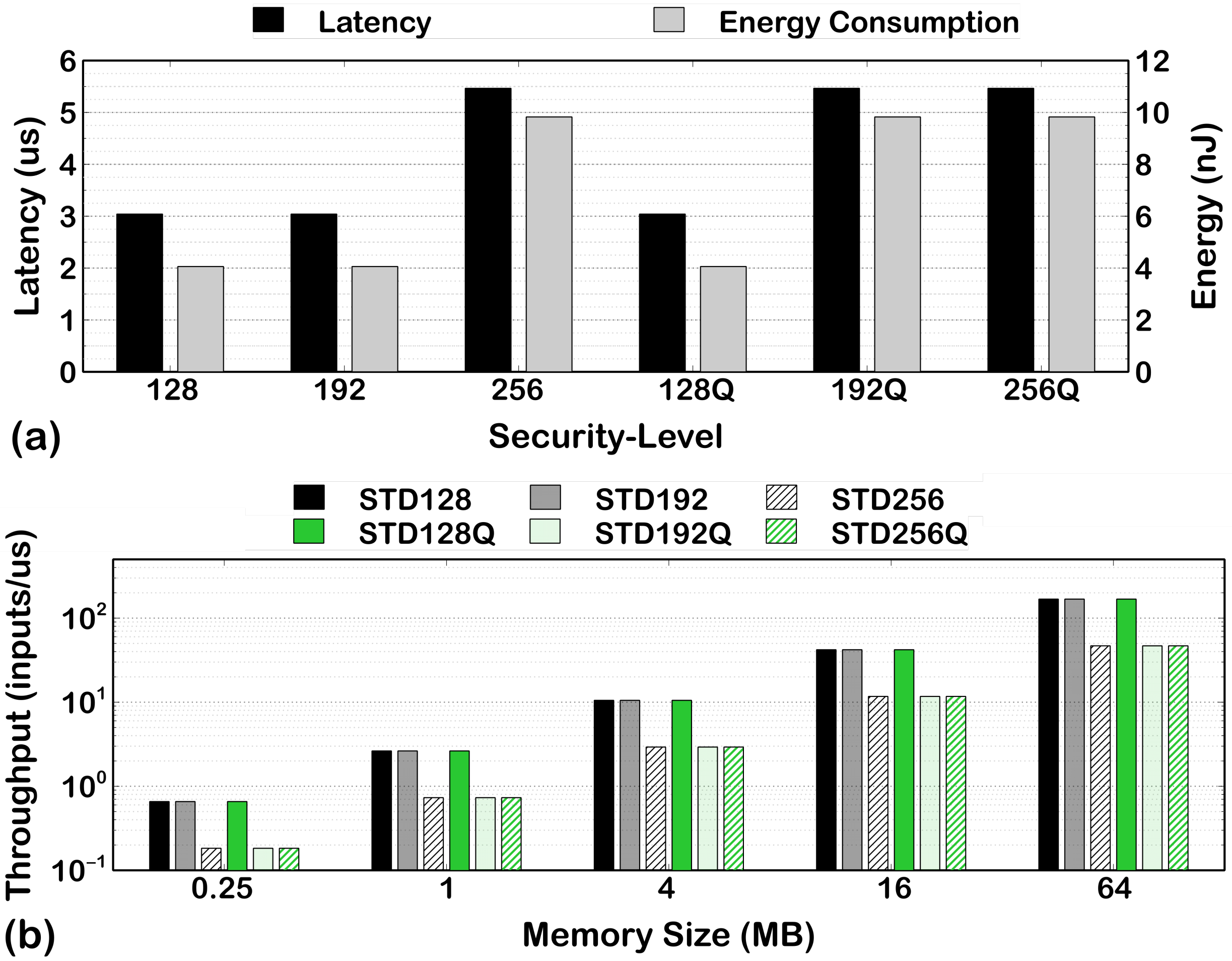

Fig. 6. Encryption in MemFHE-client. (a) Latency and energy consumption and (b) throughput for different memory sizes.

Figure 7 shows the latency of running different types of additions and multiplications in MemFHE pipeline for various security settings. We observe that for individual operations, the latency is limited by their critical path. The latencies for individual addition vary with $O(log_2 b)$, where $b$ is the bitwidth of operation, taking 353 ms (705 ms) for an 8-bit (64-bit) addition while providing 256-bit quantum-safe security. For multiplication, the latency varies with $O(b.log_2 b)$, taking 2.8 s (45 s) for an 8-bit (64-bit) multiplication.

Implementing 1024 independent additions and multiplications does not increase the latency significantly. Instead, these independent inputs fill up MemFHE's pipeline, which was otherwise severely underutilized. For example, performing 1024 8-bit additions/multiplication take only twice the total time as that for single addition/multiplication in 128-bit quantum-safe setting. For 256-bit quantum-safe FHE, the latency for 1024 8-bit additions/multiplications is actually similar to that for a single addition/multiplication. This happens because MemFHE pipeline for STD256Q is much deeper than that of STD128Q, allowing more operations to fill up the pipeline. Even for 1024 64-bit

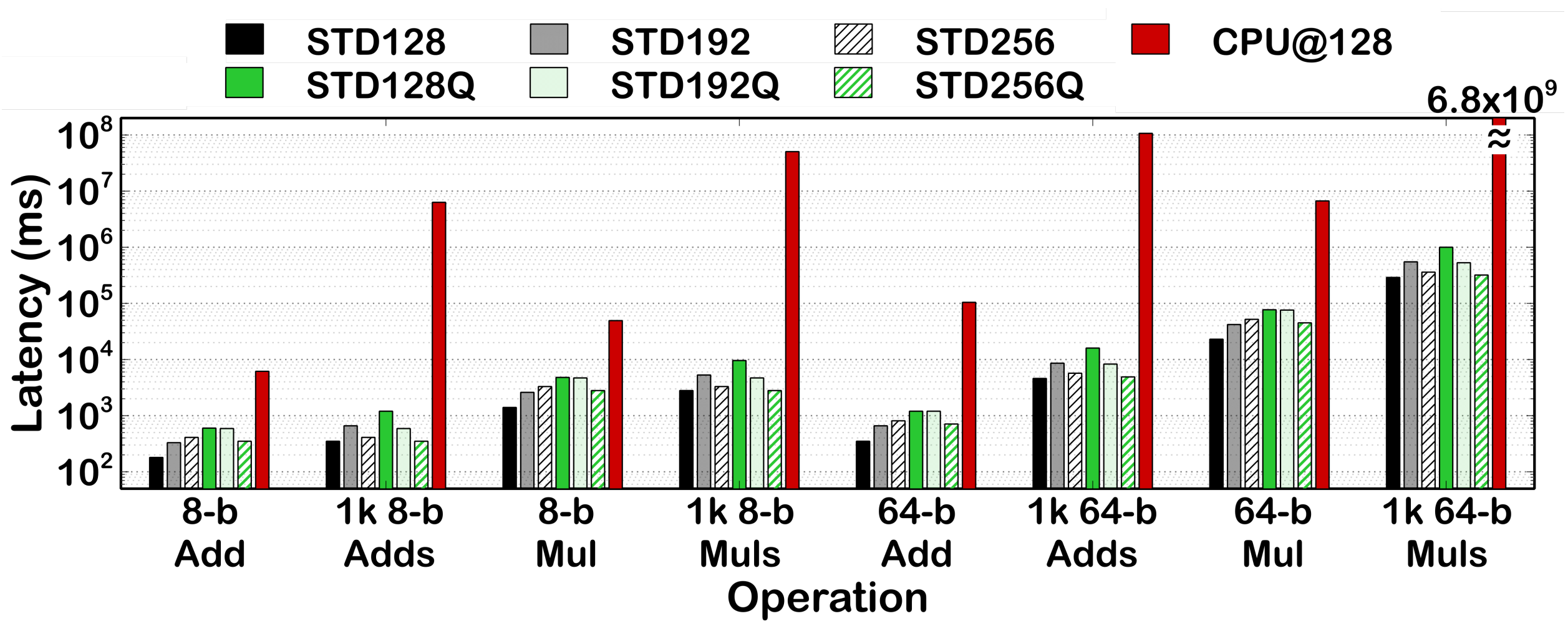

Fig. 7. End-to-end latency for implementing add and multiply ops in MemFHE.

Table 4. Workloads for Learning in MemFHE [37]

| Dataset | Network Topology | Accuracy | #GateOps |
|---|---|---|---|
| **MNIST** | C-B-A-P-C-P-F-B-A-F[20] | 99.54% | 856K |
| **CIFAR-10** | [C-B-A-C-B-A-P]×3-F-F[4] | 92.54% | 211M |
| **ImageNet** | ShuffleNet [55] | 69.4% | 1.1G |
| **Penn Treebank** [33] | LSTM: t-step 25, 300-unit layer; ReLU [33] | 89.8 PPW | 24.4M |

C: convolution layer; A: activation layer; B: batch normalization;
P: pooling layer; F: fully-connected layer; PPW: perplexity per word.

multiplications, MemFHE is at most 13× slower than one 64-bit multiplication. Hence, MemFHE truly shines when there are enough independent operations to fill the pipeline.

Lastly, Figure 7 also shows the latency of different addition and multiplication operations, normalized to MemFHE, for an Intel i7-9700 CPU with 64 GB of RAM in 128-bit classical security setting in log scale. The results were obtained using single-threaded implementation of the state-of-the-art PALISADE library [48] as detailed in [39]. We observe that CPU is on average 35× (295×) slower than MemFHE for individual 8-bit (64-bit) arithmetic operations. For 1024 arithmetic operations, MemFHE is on average 20573× faster than CPU. This is due to the highly pipelined architecture of MemFHE that can deliver higher throughput for large data. We also compare MemFHE with Nvidia GTX 1080 GPU with 8GB memory [41]. We see that MemFHE is on average 53× faster than GPU for 32-element long vector additions and multiplications. However, the latency of FHE computations in [41] scales linearly with vector-length beyond 8, while MemFHE is able to maintain the same latency for a vector-length of 160 for 32-bit multiplications. This makes MemFHE up to 265× faster than GPU.

### 9.6 Learning in MemFHE

We show MemFHE performance for complicated learning tasks. Our evaluation is inspired from the CPU implementation of TFHE-based deep neural networks (DNN) in [37], which we refer to as TDNN for simplicity. TDNN converts DNN operations into TFHE compatible functions. We use the same functions to evaluate MemFHE as it also supports TFHE. Table 4 details the datasets and the corresponding network topologies used for evaluation. TDNN works in both fully homomorphic (TDNN-FHE) mode as well as leveled mode (TDNN-Lvl). While TDNN-FHE bootstraps each gate operation, TDNN-Lvl bootstraps only higher-level operations like polynomial multiplications and additions [37].

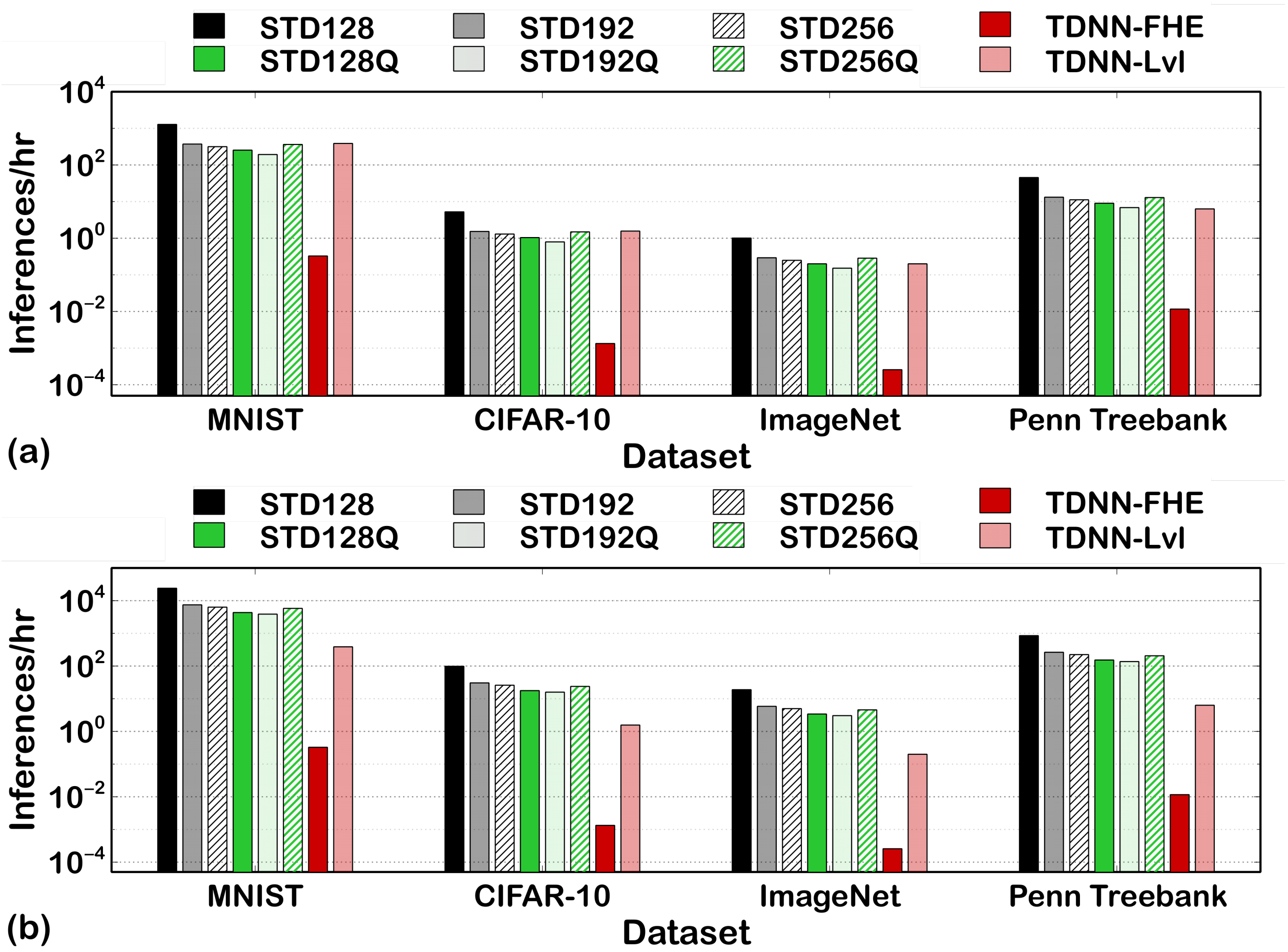

Fig. 8. Inference throughput of MemFHE and TDNN [37] for different datasets. MemFHE utilizes (a) 64GB memory and (b) 1TB memory. TDNN-FHE and TDNN-Lvl provide 163-bit and 152-bit security guarantees.

Figure 8a shows the inference throughput of MemFHE and TDNN over various datasets. MemFHE is scaled to have a total of 64GB memory size. While MemFHE provides a range of classical and quantum-safe security guarantees, TDNN provides 163-bit (152-bit) security guarantee in FHE (leveled) mode. We observe that as compared to TDNN-FHE, MemFHE provides on average 2007× higher throughput (inference/s) for classical FHE. Moreover, MemFHE has 827× higher throughput while ensuring quantum-safe FHE while TDNN-FHE just provides classical security. We also observe that MemFHE in quantum-safe provides similar throughput as TDNN-Lvl. This is a huge improvement because leveled HE accelerates computations on encrypted data by performing multiple operations without bootstrapping. However, it limits the achievable security levels. Moreover, encrypting in leveled mode is dependent on the complexity of target operation and cannot implement arbitrary operations. MemFHE achieves the throughput of a leveled implementation while running FHE.

TDNN presented in [37] runs on an Intel Xeon E7-4850 CPU with 1TB DRAM. To perform a similar memory size evaluation, we also scale MemFHE up to 1TB memory. Figure 8 summarizes the results. We observe that MemFHE's throughput further increases on average by 19× (17×) for classical (quantum-safe) FHE. This translates to four orders of magnitude higher throughput than TDNN-FHE. This huge improvement in MemFHE comes from (i) significant reduction in total data-transfers and (ii) the significantly higher number of processing in memory cores. Unlike traditional systems, off-chip data-transfers in MemFHE consists only of the communication between client and server. The high density of memory allows us to have a large number of PIM-enabled cores in the system, allowing for higher parallelism and deeper pipelining.

## 10 CONCLUSION

We presented MemFHE, the first end-to-end acceleration of fully homomorphic encryption in PIM. We designed accelerators for both client as well as server for the latest RGSW based homomorphic encryption schemes. MemFHE reduces the data transfer bottlenecks and enables extensive parallelism. MemFHE raises the bar of today's systems security, providing both classical and quantum-safe security guarantees.

## ACKNOWLEDGMENTS

This work was supported in part by CRISP, one of six centers in JUMP, an SRC program sponsored by DARPA, in part by SRC-Global Research Collaboration grants, and also NSF grants # 1527034, #1730158, #1826967, #1911095, and #2003279.

## REFERENCES

[1] [n.d.]. NuFHE, a GPU-powered Torus FHE implementation. *Available at https://github.com/nucypher/nufhe* ([n. d.]).

[2] Jacob Alperin-Sheriff and Chris Peikert. 2014. Faster bootstrapping with polynomial error. In *Annual Cryptology Conference*. Springer, 297–314.

[3] Paul Barrett. 1986. Implementing the Rivest Shamir and Adleman Public Key Encryption Algorithm on a Standard Digital Signal Processor. In *CRYPTO*.

[4] Hervé Chabanne, Amaury de Wargny, Jonathan Milgram, Constance Morel, and Emmanuel Prouff. 2017. Privacy-Preserving Classification on Deep Neural Network. *IACR Cryptol. ePrint Arch.* 2017 (2017), 35.

[5] Hao Chen and Kyoohyung Han. 2018. Homomorphic lower digits removal and improved FHE bootstrapping. In *Annual International Conference on the Theory and Applications of Cryptographic Techniques*. Springer, 315–337.

[6] Hao Chen, Zhicong Huang, Kim Laine, and Peter Rindal. 2018. Labeled PSI from fully homomorphic encryption with malicious security. In *Proceedings of the 2018 ACM SIGSAC Conference on Computer and Communications Security*. 1223–1237.

[7] Ilaria Chillotti, Nicolas Gama, Mariya Georgieva, and Malika Izabachene. 2016. Faster fully homomorphic encryption: Bootstrapping in less than 0.1 seconds. In *international conference on the theory and application of cryptology and information security*. Springer, 3–33.

[8] Ilaria Chillotti, Nicolas Gama, Mariya Georgieva, and Malika Izabachène. 2020. TFHE: fast fully homomorphic encryption over the torus. *Journal of Cryptology* 33, 1 (2020), 34–91.

[9] Edward J Chou, Arun Gururajan, Kim Laine, Nitin Kumar Goel, Anna Bertiger, and Jack W Stokes. 2020. Privacy-preserving phishing web page classification via fully homomorphic encryption. In *ICASSP 2020-2020 IEEE International Conference on Acoustics, Speech and Signal Processing (ICASSP)*. IEEE, 2792–2796.

[10] Hüsrev Cılasun, Salonik Resch, Zamshed Iqbal Chowdhury, Erin Olson, Masoud Zabihi, Zhengyang Zhao, Thomas Peterson, Jian-Ping Wang, Sachin S Sapatnekar, and Ulya Karpuzcu. 2020. Crafft: High resolution fft accelerator in spintronic computational ram. In *2020 57th ACM/IEEE Design Automation Conference (DAC)*. IEEE, 1–6.

[11] David Bruce Cousins, Kurt Rohloff, and Daniel Sumorok. 2016. Designing an FPGA-accelerated homomorphic encryption co-processor. *IEEE Transactions on Emerging Topics in Computing* 5, 2 (2016), 193–206.

[12] Wei Dai and Berk Sunar. [n.d.]. Cuda-accelerated fully homomorphic encryption library, August 2019.

[13] Léo Ducas and Daniele Micciancio. 2015. FHEW: bootstrapping homomorphic encryption in less than a second. In *Annual International Conference on the Theory and Applications of Cryptographic Techniques*. Springer, 617–640.

[14] Charles Eckert, Xiaowei Wang, Jingcheng Wang, Arun Subramaniyan, Ravi Iyer, Dennis Sylvester, David Blaaauw, and Reetuparna Das. 2018. Neural cache: Bit-serial in-cache acceleration of deep neural networks. In *2018 ACM/IEEE 45th Annual International Symposium on Computer Architecture (ISCA)*. IEEE, 383–396.

[15] Daichi Fujiki, Scott Mahlke, and Reetuparna Das. 2019. Duality cache for data parallel acceleration. In *Proceedings of the 46th International Symposium on Computer Architecture*. 397–410.

[16] Nicolas Gama, Malika Izabachene, Phong Q Nguyen, and Xiang Xie. 2016. Structural lattice reduction: generalized worst-case to average-case reductions and homomorphic cryptosystems. In *Annual International Conference on the Theory and Applications of Cryptographic Techniques*. Springer, 528–558.

[17] Craig Gentry. 2009. Fully homomorphic encryption using ideal lattices. In *Proceedings of the forty-first annual ACM symposium on Theory of computing*. 169–178.

[18] Craig Gentry, Amit Sahai, and Brent Waters. 2013. Homomorphic encryption from learning with errors: Conceptually-simpler, asymptotically-faster, attribute-based. In *Annual Cryptology Conference*. Springer, 75–92.

[19] Amirali Ghofrani, Abbas Rahimi, Miguel A Lastras-Montaño, Luca Benini, Rajesh K Gupta, and Kwang-Ting Cheng. 2016. Associative memristive memory for approximate computing in gpus. *IEEE Journal on Emerging and Selected Topics in Circuits and Systems* 6, 2 (2016), 222–234.

[20] Ran Gilad-Bachrach, Nathan Dowlin, Kim Laine, Kristin Lauter, Michael Naehrig, and John Wernsing. 2016. Cryptonets: Applying neural networks to encrypted data with high throughput and accuracy. In *International Conference on Machine Learning*. PMLR, 201–210.

[21] Alvin Oliver Glova, Itir Akgun, Shuangchen Li, Xing Hu, and Yuan Xie. 2019. Near-data acceleration of privacy-preserving biomarker search with 3D-stacked memory. In *2019 Design, Automation & Test in Europe Conference & Exhibition (DATE)*. IEEE, 800–805.

[22] Antonio Guimarães, Edson Borin, and Diego F Aranha. 2021. Revisiting the functional bootstrap in TFHE. *IACR Transactions on Cryptographic Hardware and Embedded Systems* (2021), 229–253.

[23] Saransh Gupta, Mohsen Imani, and Tajana Rosing. 2018. FELIX: Fast and Energy-Efficient Logic in Memory. In *Proceedings of the International Conference on Computer-Aided Design*. ACM, 55.

[24] Ameer Haj-Ali, Rotem Ben-Hur, Nimrod Wald, and Shahar Kvatinsky. 2018. Efficient algorithms for in-memory fixed point multiplication using magic. In *2018 IEEE International Symposium on Circuits and Systems (ISCAS)*. IEEE, 1–5.

[25] Ameer Haj-Ali, Rotem Ben-Hur, Nimrod Wald, Ronny Ronen, and Shahar Kvatinsky. 2018. Imaging: In-memory algorithms for image processing. *IEEE Transactions on Circuits and Systems I: Regular Papers* 65, 12 (2018), 4258–4271.

[26] Shai Halevi and Victor Shoup. 2014. Algorithms in helib. In *Annual Cryptology Conference*. Springer, 554–571.

[27] Mohsen Imani, Saransh Gupta, Yeseong Kim, and Tajana Rosing. 2019. Floatpim: In-memory acceleration of deep neural network training with high precision. In *2019 ACM/IEEE 46th Annual International Symposium on Computer Architecture (ISCA)*. IEEE, 802–815.

[28] Mohsen Imani, Saikishan Pampana, Saransh Gupta, Minxuan Zhou, , Yeseong Kim, and Tajana Rosing. 2020. DUAL: Acceleration of Clustering Algorithms using Digital-based Processing In-Memory. In *Proceedings of the International Symposium on Microarchitecture*. IEE/ACM.

[29] Miran Kim, Arif Harmanci, Jean-Philippe Bossuat, Sergiu Carpov, Jung Hee Cheon, Ilaria Chillotti, Wonhee Cho, David Froelicher, Nicolas Gama, Mariya Georgieva, et al. 2020. Ultra-Fast Homomorphic Encryption Models enable Secure Outsourcing of Genotype Imputation. *bioRxiv* (2020).

[30] Miran Kim, Yongsoo Song, Baiyu Li, and Daniele Micciancio. 2020. Semi-parallel logistic regression for GWAS on encrypted data. *BMC Medical Genomics* 13, 7 (2020), 1–13.

[31] Shahar Kvatinsky, Dmitry Belousov, Slavik Liman, Guy Satat, Nimrod Wald, Eby G Friedman, Avinoam Kolodny, and Uri C Weiser. 2014. MAGIC – Memristor-aided logic. *IEEE Transactions on Circuits and Systems II: Express Briefs* 61, 11 (2014), 895–899.

[32] Shahar Kvatinsky, Misbah Ramadan, Eby G Friedman, and Avinoam Kolodny. 2015. VTEAM: A general model for voltage-controlled memristors. *IEEE Transactions on Circuits and Systems II: Express Briefs* 62, 8 (2015), 786–790.

[33] Quoc V Le, Navdeep Jaitly, and Geoffrey E Hinton. 2015. A simple way to initialize recurrent networks of rectified linear units. *arXiv preprint arXiv:1504.00941* (2015).

[34] Moon Sung Lee, Yongje Lee, Jung Hee Cheon, and Yunheung Paek. 2015. Accelerating bootstrapping in FHEW using GPUs. In *2015 IEEE 26th International Conference on Application-specific Systems, Architectures and Processors (ASAP)*. IEEE, 128–135.

[35] Xinya Lei, Ruixin Guo, Feng Zhang, Lizhe Wang, Rui Xu, and Guangzhi Qu. 2019. Optimizing FHEW With Heterogeneous High-Performance Computing. *IEEE Transactions on Industrial Informatics* 16, 8 (2019), 5335–5344.

[36] Zhenyu Liu, Yang Song, Takeshi Ikenaga, and Satoshi Goto. 2005. A VLSI array processing oriented fast Fourier transform algorithm and hardware implementation. *IEICE Transactions on Fundamentals of Electronics, Communications and Computer Sciences* 88, 12 (2005), 3523–3530.

[37] Qian Lou and Lei Jiang. 2019. SHE: A Fast and Accurate Deep Neural Network for Encrypted Data. *Advances in neural information processing systems* (2019).

[38] Vadim Lyubashevsky, Chris Peikert, and Oded Regev. 2010. On ideal lattices and learning with errors over rings. In *Annual International Conference on the Theory and Applications of Cryptographic Techniques*. Springer, 1–23.

[39] Daniele Micciancio and Yuriy Polyakov. 2020. Bootstrapping in FHEW-like Cryptosystems. *IACR Cryptol. ePrint Arch.* 2020 (2020), 86.

[40] Peter L Montgomery. 1985. Modular Multiplication Without Trial Division. *Mathematics of computation* (1985).

[41] Toufique Morshed, Md Momin Al Aziz, and Noman Mohammed. 2020. CPU and GPU Accelerated Fully Homomorphic Encryption. In *2020 IEEE International Symposium on Hardware Oriented Security and Trust (HOST)*. IEEE, 142–153.

[42] Hamid Nejatollahi, Saransh Gupta, Mohsen Imani, Tajana Simunic Rosing, Rosario Cammarota, and Nikil Dutt. 2020. CryptoPIM: in-memory acceleration for lattice-based cryptographic hardware. In *2020 57th ACM/IEEE Design Automation Conference (DAC)*. IEEE, 1–6.

[43] Dimin Niu, Qiaosha Zou, Cong Xu, and Yuan Xie. 2013. Low power multi-level-cell resistive memory design with incomplete data mapping. In *2013 IEEE 31st International Conference on Computer Design (ICCD)*. IEEE, 131–137.

[44] Oded Regev. 2009. On lattices, learning with errors, random linear codes, and cryptography. *Journal of the ACM (JACM)* 56, 6 (2009), 1–40.

[45] Dayane Reis, Michael T Niemier, and Xiaobo Sharon Hu. 2019. A computing-in-memory engine for searching on homomorphically encrypted data. *IEEE Journal on Exploratory Solid-State Computational Devices and Circuits* 5, 2 (2019), 123–131.

[46] Dayane Reis, Jonathan Takeshita, Taeho Jung, Michael Niemier, and Xiaobo Sharon Hu. 2020. Computing-in-Memory for Performance and Energy-Efficient Homomorphic Encryption. *IEEE Transactions on Very Large Scale Integration (VLSI) Systems* 28, 11 (2020), 2300–2313.

[47] M Sadegh Riazi, Kim Laine, Blake Pelton, and Wei Dai. 2020. HEAX: An architecture for computing on encrypted data. In *Proceedings of the Twenty-Fifth International Conference on Architectural Support for Programming Languages and Operating Systems*. 1295–1309.

[48] Kurt Rohloff and Yuriy Polyakov. [n.d.]. The PALISADE Lattice Cryptography Library, 1.2017. *Library available at https://git. njit. edu/palisade/PALISADE* ([n. d.]).

[49] Sujoy Sinha Roy, Furkan Turan, Kimmo Jarvinen, Frederik Vercauteren, and Ingrid Verbauwhede. 2019. FPGA-based high-performance parallel architecture for homomorphic computing on encrypted data. In *2019 IEEE International symposium on high performance computer architecture (HPCA)*. IEEE, 387–398.

[50] Nikola Samardzic, Axel Feldmann, Aleksandar Krastev, Srinivas Devadas, Ronald Dreslinski, Christopher Peikert, and Daniel Sanchez. 2021. F1: A Fast and Programmable Accelerator for Fully Homomorphic Encryption. In *MICRO-54: 54th Annual IEEE/ACM International Symposium on Microarchitecture*. 238–252.

[51] R Singleton. 1967. A method for computing the fast Fourier transform with auxiliary memory and limited high-speed storage. *IEEE Transactions on Audio and Electroacoustics* 15, 2 (1967), 91–98.

[52] Nishil Talati, Saransh Gupta, Pravin Mane, and Shahar Kvatinsky. 2016. Logic design within memristive memories using memristor-aided loGIC (MAGIC). *IEEE Transactions on Nanotechnology* 15, 4 (2016), 635–650.

[53] J Joshua Yang, Dmitri B Strukov, and Duncan R Stewart. 2013. Memristive devices for computing. *Nature nanotechnology* 8, 1 (2013), 13–24.

[54] Hasan Erdem Yantir, Wenzhe Guo, Ahmed M Eltawil, Fadi J Kurdahi, and Khaled Nabil Salama. 2019. An ultra-area-efficient 1024-point in-memory fft processor. *Micromachines* 10, 8 (2019), 509.

[55] Xiangyu Zhang, Xinyu Zhou, Mengxiao Lin, and Jian Sun. 2018. Shufflenet: An extremely efficient convolutional neural network for mobile devices. In *Proceedings of the IEEE conference on computer vision and pattern recognition*. 6848–6856.

[56] Junwei Zhou, Junjiong Li, Emmanouil Panaousis, and Kaitai Liang. 2020. Deep Binarized Convolutional Neural Network Inferences over Encrypted Data. In *2020 7th IEEE International Conference on Cyber Security and Cloud Computing (CSCloud)/2020 6th IEEE International Conference on Edge Computing and Scalable Cloud (EdgeCom)*. IEEE, 160–167.